\documentclass[12pt,a4paper]{article}
\title{A Discrete Phase-Space Calculus for Quantum Spins based on a Reconstruction Method using Coherent States\footnote{Contribution to the Sixth Central-European Workshop on Quantum Optics, Chateau Chudobin near Olomouc, April 30 - May 3, 1999}}
\author{Stefan Weigert \\
Institut de Physique, Universit\'e de Neuch\^atel\\
Rue A.-L. Breguet 1, CH-2000 Neuch\^atel, Switzerland\\
\texttt{stefan.weigert@iph.unine.ch}}
\date{April 1999}
\newcommand\be{\begin{equation}}
\newcommand\ee{\end{equation}}
\newcommand\ket[1]{|#1\rangle}
\newcommand\bra[1]{\langle #1|}

\begin{document}
\maketitle
\begin{abstract}
To reconstruct a mixed or pure quantum state of a spin $s$ is possible through  coherent states: its density matrix is fixed by the probabilities to measure the value $s$ along $4s(s+1)$ appropriately chosen directions in space. Thus, after inverting the experimental data, the statistical operator is parametrized entirely by expectation values. On this basis, a symbolic calculus for quantum spins is developed, the `expectation-value representation.' It resembles the Moyal representation for $SU(2)$ but two important differences exist. On the one hand, the symbols take values on a discrete set of points in phase space only. On the other hand, no quasi-probabilities---that is, phase-space distributions with {\em negative} values---are encountered in this approach.
\end{abstract}
\section{Introduction}
Coherent states provide a versatile tool in quantum mechanics both from a conceptual and a technical point of view. Originally discovered \cite{schroedinger26} in the search for `particle solutions'  (stable, nondispersing wave packets) of Schr\"odinger's equation, they turned out to be useful \cite{perelomov86}, for example, in semiclassical descriptions of quantum systems, quantum optics, quantum statistical mechanics, quantum field theory$\ldots$ To a large extent, this wide range of applications  is due to the intimate link of coherent states to group theory. Having once realized this fundamental connection, the way is open to generalize the initial version of coherent states which is related to the Heisenberg-Weyl group of phase-space translations of a quantum particle.
 
The purpose of the present note is twofold. First, it will be pointed out that coherent states enable one to provide a simple answer to the question: {\em How to express the state of a quantum mechanical system by measurable quantities?} Due to recent experimental progress in handling individual quantum systems \cite{smithey+93}-\cite{kurtsiefer+97}, this long-standing question \cite{pauli33} has turned into a rapidly expanding field known as {\em state reconstruction} \cite{leonhardt97}. 

Second, a {\em discrete phase-space formulation} of the quantum mechanics will be shown to emerge naturally from the coherent-state approach to state reconstruction for a spin. The new `expectation value representation' deals with {\em positive} probabilities in phase space only, contrary to the Wigner functions or `quasi-probabilities' which may take negative values, too. The resulting symbolic calculus provides a {\em generalized} Stratonovich-Weyl correspondence between 
operators and symbols  as follows from a comparison to the Moyal representation for a spin.            

There is a large variety of other methods to determine the state of a spin in terms of expectations. Many references can be found in Refs.\  \cite{weigert92}-\cite{weigert99/1}, and  \cite{brif+99}, for example.
\section{State reconstruction based on coherent states}
Let us briefly review the quantum mechanical description of a spin which is based on a vector operator $\widehat {\bf S} \equiv \hbar \widehat{\bf s}$. Its components act irreducibly in a $(2s+1)$-dimensional Hilbert space ${\cal H}_s$, and they satisfy the commutation relations of the algebra $su(2)$: $[ {\hat s}_x, {\hat s}_y ]= i{\hat s}_z, \ldots$ The standard basis of the space ${\cal H}_s$ is given by the eigenvectors of the $z$ component  of the spin, ${\widehat S}_z =  \widehat {\bf S} \cdot {\bf n}_z$, denoted by $\ket{\mu, {\bf n}_z},$ $-s \leq \mu \leq s$. Observables are represented by hermitian operators, ${\widehat {A}}^\dagger = {\widehat {A}}$, all of which are linear combinations of polynomials in the operators ${\hat s}_x$, ${\hat s}_y$ and ${\hat s}_z$ of degree $2s$ at most. The ensemble of all hermitean operators acting on ${\cal H}_s$ can be considered as a vector space ${\cal A}_s$ of dimension $N_s = (2s+1)^2$.

A {\em coherent spin state} $\ket{ {\bf n}}$ is associated to each  point ${\bf n} = ( \sin \vartheta \cos \varphi, \sin \vartheta  \sin \varphi,$ $\cos \vartheta) \in {\cal S}^2$ of the surface of the unit sphere \cite{arecchi+72}. It coincides with the eigenstate of the operator $ \hat {\bf s} \cdot {\bf n} $ along the direction ${\bf n} $ and with eigenvalue $s$:
\be
\ket{{\bf n}} 
        \equiv  \exp [ -i \, \vartheta \, {\bf m}(\varphi) \cdot {\hat {\bf s}} \, ] \, \ket{s,{\bf n}_z} \, ,
\label{defineaxes}
\ee
where ${\bf m}(\varphi) = (- \sin \varphi,\cos\varphi,0)$. Thus, the coherent state $\ket{{\bf n} }$ is obtained from rotating the state $\ket{s,{\bf n}_z}$ about the axis ${\bf m}(\varphi)$ in the $xy$ plane by an angle $\vartheta$. The ensemble of all coherent states provides an overcomplete basis of the Hilbert space ${\cal H}_s$:
\be
\frac{(2s+1)}{4\pi} \int_{{\cal S}^2} d {\bf n} \, \ket{{\bf n}} \bra{{\bf n}} 
= {\widehat E} \, ,
\label{complete}
\ee
where ${\widehat E}$ is the unit operator on ${\cal H}_s$.
It is convenient to combine $(\vartheta,\varphi)$ into a single complex variable, $z= \tan(\vartheta/2)\exp [ i \varphi ]$. This provides a stereographic projection of the surface of the sphere to the complex plane. In terms of $z$, a coherent state has the expansion  \cite{amiet+91}
\be
\ket{ z }  
     = \frac{1}{(1+|z|^2)^s}\sum_{k= 0}^{2s}
           \left(
                  \begin{array}{c}
                      2s \\ 
                       k 
                  \end{array}
           \right)^{1/2} 
z^{k}\ket{s-k, {\bf n}_z} \, .
\label{expandcs}
\ee

In order to show that the density matrix ${\hat \rho}$ of a spin $s$ is determined unambiguously by appropriate measurements with a Stern-Gerlach apparatus one preceeds as follows. Distribute $N_s = (2s+1)^2$ axes ${\bf n}_{\mu\nu}, -s \leq \mu,\nu \leq s$, over $(2s+1)$ cones about the $z$ axis with different opening angles such that the set of the $(2s+1)$ directions on each cone is invariant under a rotation about $z$ by an angle $2\pi/(2s+1)$. As shown in  \cite{amiet+99/2}, an unnormalized statistical operator ${\hat \rho}$ is then fixed by measuring the $N_s$ relative frequencies 
\be
p_s({\bf n}_{\mu\nu}) = \bra{ {\bf n}_{\mu\nu}} \hat \rho \ket{ {\bf n}_{\mu\nu}} \, ,
 \qquad  -s \leq \mu,\nu \leq s \, , 
\label{relfreq}
\ee
that is, by the expectation values of the statistical operator $\hat \rho$ in the coherent states $\ket{{\bf n}_{\mu\nu}}$. Here is the idea of the proof: take
the expectation value of $\hat \rho$ in the coherent states $\ket{ {\bf n}_{\mu\nu}}$
as given in Eq. (\ref{expandcs}). You obtain $N_s$ linear relations between probabilities 
$p_s({\bf n}_{\mu\nu})$ and the matrix elements of the density matrix with respect to the basis $\ket{s-k, {\bf n}_z}$. This set of equations can be inverted by standard techniques if the directions ${\bf n}_{\mu\nu}$ are chosen as described above. 

For a spin $s$, the projection operators 
\be
{\widehat Q}_{{\mu\nu}} 
        = \ket{{\bf n}_{{\mu\nu}}} \bra{{\bf n}_{{\mu\nu}}}   \, , 
          \qquad  -s \leq \mu,\nu \leq s  \, ,
\label{mixedrho}
\ee
constitute thus a {\em quorum} ${\cal Q}$. In general, a quorum is defined as a collection of (hermitean) operators having the property that their expectation values are sufficient to reconstruct the quantum state of the system at hand. The operators 
${\widehat Q}_{{\mu\nu}}$ do even define an {\em optimal} quorum since exactly $(2s+1)^2$ numbers have to be determined experimentally which equals the number of free real parameters of the (unnormalized) hermitean density matrix $\hat \rho$.
 
If the state to be reconstructed is known to be a {\em pure} one, only $4s$ real parameters need to be specified in order to identify it. The set of operators ${\widehat Q}_{{\mu\nu}} $ still provides a quorum but the data required for reconstruction now is highly redundant. In Ref.\ \cite{amiet+99/2}, another reconstruction method based on coherent states has been worked out which takes into account the reduced number of parameters.    
\section{A symbolic calculus for spin systems}
It will be argued now that Eqs.\ (\ref{relfreq}) and (\ref{mixedrho}) provide the basis for a {\em symbolic calculus} in the spirit of the Wigner formalism  \cite{wigner32,amiet+91} of quantum mechanics. For simplicity, the labels $\mu$ and $\nu$ are replaced by a single index $n= (\mu,\nu)$, say, with $1 \leq n \leq N_s$.  

The set of all unnormalized hermitean density matrices for a spin $s$ is equal to the set of all hermitean operators acting on the Hilbert space ${\cal H}_s$. Therefore, the expectation values 
\begin{equation}
{\sf A}_{n}  = \mbox{ Tr } \left[ {\widehat {A}} \, {\widehat Q}_{n} \right] \, ,
\qquad 1 \leq n \leq N_s \, ,
\label{symbol1}
\end{equation}
of a hermitean operator $\widehat{A}$ in the coherent states $\ket{{\bf n}_n}$ 
suffice to characterize unambiguously the operator. In other words, the is a one-to-one relation between the operator $\widehat{A}$ and its {\em symbol\ }, the collection of $N_s$ real numbers ${\sf A}_{n}$. Considering the trace as a scalar product, Eq. (\ref{symbol1}) can be interpreted as the projection of the operator $\widehat{A}$ on $N_s$ linearly independent elements ${\widehat Q}_{n}$ of the vector space ${\cal A}_s$. The operators ${\widehat Q}_{n}$, however, do {\em not} constitute an orthogonal basis: $\mbox{ Tr } [ {\widehat Q}_{n}{\widehat Q}_{n'} ] = | \langle {\bf n}_n \ket{{\bf n}_{n'}} |^2 \neq 0$. Nevertheless, the inversion of Eq. (\ref{symbol1}) is achieved easily by means of the operators ${\widehat {\sf Q}}^{n}$, a second basis of the space ${\cal A}_s$, defined as the dual of the quorum (\ref{mixedrho}): 
\begin{equation}
\frac{1}{2s+1} \mbox{ Tr }\left[  {\widehat Q}_{n} {\widehat {\sf Q}}^{n'} \right] 
= \delta_{n}^{n'}  \, , 
\qquad 1 \leq n,n' \leq N_s \, .
\label{orthogonalityquorum}
\end{equation}
This implies an expansion for hermitean operators,
\begin{equation}
{\widehat {A}} = \frac{1}{2s+1} \sum_{n=1}^{N_s} {\sf A}_{n} {\widehat {\sf Q}}^{n} \, ,
\label{expandA}
\end{equation}
which is indeed the inversion of (\ref{symbol1}): ${\widehat {A}}$ is given explicitly as a function of the measurable numbers ${\sf A}_{n}$ and a specified set of operators. Due to the symmetry between the quorum ${\cal Q}$ and its dual ${\cal Q}^D$, there is a second way to expand hermitean operators,  
\begin{equation}
{\widehat {A}} = \frac{1}{2s+1} \sum_{n=1}^{N_s} A^{n} {\widehat Q}_{n} \, ,
\qquad 
A^{n}  = \mbox{ Tr } \left[ {\widehat {A}} \, {\widehat {\sf Q}}^{n} \right] \, ,
\label{expandquorum}
\end{equation}
with coefficients $A^n \neq {\sf A}_{n}$, providing a {\em second} symbol of ${\widehat {A}}$. 

Upon expanding the operator ${\widehat Q}_n$ in terms of the dual basis, 
\be
{\widehat Q}_n = \frac{1}{2s+1} \sum_{n'=1}^{N_s} {\sf G}_{nn'} {\widehat {\sf Q}}^{n'} \, , 
\qquad {\sf G}_{nn'} = \mbox{ Tr } \left[ {\widehat Q}_{n} {\widehat Q}_{n'} \right] \, , 
\label{trfqdualq}
\ee
one discovers the existence of a real symmetric matrix ${\sf G}$ which can be shown to be {\em positive definite}. It can be used as a {\em metric} for raising and lowering indeces of both symbols and operators.

In terms of symbols, tracing the product ${\widehat {A}} \, {\widehat {B}}$ of two hermitean operators gives
\be
\mbox{ Tr } \left[  {\widehat {A}}\,  {\widehat {B}} \right]
   = \frac{1}{(2s+1)} \sum_{n=1}^{N_s} {\widehat {\sf A}}^{n} {\widehat B}_{n}
   = \frac{1}{(2s+1)} \sum_{n=1}^{N_s} {\widehat A}_{n} {\widehat {\sf B}}^{n} \, ,
\label{symbprod}
\ee
using the expansions (\ref{expandA}), (\ref{expandquorum}), and (\ref{orthogonalityquorum}). Invoking the metric ${\sf G}$, one can also write
\be
\mbox{ Tr } \left[  {\widehat {A}}\,  {\widehat {B}} \right]
   = \frac{1}{(2s+1)^2} \sum_{n,n'=1}^{N_s} 
             {\sf G}_{nn'} {\widehat {\sf A}}^{n} {\widehat {\sf B}}^{n'}
   = \frac{1}{(2s+1)^2} \sum_{n,n'=1}^{N_s} 
              {\sf G}^{nn'} {\widehat A}_{n} {\widehat B}_{n'} \, .
\label{symbprod2}
\ee

Let us now consider the properties of the statistical operator $\hat \rho$  
when expanded in the basis ${\widehat {\sf Q}}^{n}$ dual to the original quorum,
\begin{equation}
{\hat {\rho}} = 
    \frac{1}{2s+1} \sum_{n=1}^{N_s} {\sf P}_{n} {\widehat {\sf Q}}^{n} \, ,
\label{expandrhodual}
\end{equation}
where the coefficients ${\sf P}_{n} = \mbox{ Tr } [ \hat \rho\,  {\widehat Q}_{n} ] \equiv  \bra{{\bf n}_{n}} \hat \rho \ket{{\bf n}_{n}}$ satisfy 
\be
0   \leq    {\sf P}_{n}  \leq 1 \, , 
\qquad 1 \leq n \leq N_s \, .
\label{positive!}
\ee
Since the density matrix $\hat \rho$ is a positive operator, the ${\sf P}_{n}$ are {\em non-negative} throughout, and each of the $N_s$ numbers ${\sf P}_{n}$ has a value less or equal to one due to the normalization condition $\mbox{ Tr }[\,  \hat \rho \, ] = 1 $. The positivity of the numbers ${\sf P}_{n}$ is characteristic for the basis $ {\widehat {\sf Q}}^{n} $, and it is {\em  not} valid for the expansion coefficients of $\hat \rho$ with respect to the basis ${\widehat Q}_{n}$. The interpretation of the coefficients $ {\sf P}_{n}$---to measure the value $s$ along the axis ${\bf n}_{n}$---is clearly compatible with (\ref{positive!}). 

It is important to note that, although each of the 
 $ {\sf P}_{n}$ {\em is} a probability, they do {\em not} sum up to unity:  
\be
0 < \sum_{n=1}^{N_s} {\sf P}_{n} < (2s+1)^2 \, .
\label{largersum}
\ee
This is due to the fact that they all refer to {\em different orientations} of the Stern-Gerlach apparatus, being thus associated with the measurement of {\em incompatible} observables, 
\be
\left[ {\widehat Q}_{n}, {\widehat Q}_{n'} \right] 
          \neq 0 \, ,
\qquad 1 \leq n,n' \leq N_s \, ,
\label{commutator}
\ee
since the scalar product $\bra{{\bf n}_{n}} {\bf n}_{n'} \rangle$ of two coherent states is different from zero. The sum in (\ref{largersum}) cannot take the value $(2s+1)^2 $ since this would require a common eigenstate of all the operators ${\widehat Q}_{n}$ which does not exist due to 
(\ref{commutator}). By an appropriate choice of the directions ${\bf n}_{n}$ (all in the neighborhood of one single direction ${\bf n}_0$, say), the sum
can be arbitrarily close to $(2s+1)^2 $ for states `peaked' about ${\bf n}_0$. Similarly, the sum of all ${\sf P}_{n}$ cannot take on the value zero since this would require a vanishing density matrix which is impossible. If, however, considered as a sum of {\em expectation values}, there is no need for the numbers ${\sf P}_{n}$ to sum up to unity. Nevertheless, they are not completely independent when arising from a statistical operator: its normalization  implies that 
\be
 \mbox{ Tr } \left[ \, \hat \rho \, \right]  
  = \frac{1}{2s+1} \sum_{n=1}^{N_s} 
        \mbox{ Tr} \left[ {\widehat {\sf Q}}^{n} \right] {\sf P}_{n} 
  = 1 \, ,
\label{restriction}
\end{equation}
turning one of the probabilities into a function of the $(2s+1)^2-1 = 4s(s+1)$ others, leaving us with the correct number of free real parameters needed to specify a density matrix. In Ref.\
\cite{weigert99/2}, the time evolution of the density matrix, generated by a Hamiltonian $\widehat H$, is expressed in terms of linear differential equations of first-order which couple the probabilities ${\sf P}_{n}$. The resulting `expectation-value representation' for 
a spin is equivalent to any other representation.  
\section{Comparison to the Moyal representation}
In the following, the symbolic representation introduced above is compared to the Moyal representation associated with the group $SU(2)$ as given by V\'arilly and Gracia-Bond\'\i a \cite{varilly+89}. For a spin $s$, they define a `Stratonovich-Weyl' correspondence as a rule which maps each operator ${\widehat A}$ on the 
$(2s+1)$-dimensional Hilbert space ${\cal H}_s$ to a function $W_{\! \! A} ({\bf n})$ on the phase space of the classical spin, ${\cal S}^2$. This prescription must satisfy various properties which are conveniently expressed in terms of a family of operator kernels ${\widehat \Delta ({\bf n})}$. The kernels establish a correspondence between  an operator ${\widehat A}$ and its symbol via
\be 
{W_{\! \! A} ({\bf n})} = \mbox{ Tr } \left[ {\widehat A} \, {\widehat \Delta ({\bf n})} \right] \, .
\label{definesymbol}
\ee
Here are the requirements to be satisfied by the kernels:
\begin{enumerate}
\item[($\alpha$):]{Each kernel ${\widehat \Delta} ({\bf n})$ is a hermitean operator,
\be
{\widehat \Delta}^\dagger ({\bf n}) = {\widehat \Delta} ({\bf n}) \, , \qquad
                           {\bf n} \in {\cal S}^2 \, , 
\label{herm}
\ee
implying that the symbol ${W_{\! \! A} ({\bf n})}$ in (\ref{definesymbol}) of a hermitean operator ${\widehat A}$ is {\em real}.}
\item[($\beta$):]{The set of all kernels is (over-) {\em complete}:
\be
\frac{(2s+1)}{4\pi} \int_{{\cal S}^2} d{\bf n} \,  {\widehat \Delta ({\bf n})} = {\widehat E} \, ,
\label{compl}
\ee}
providing thus an explicit resolution of the identity.
\item[($\gamma$):]{Two kernels with labels ${\bf n}$ and ${\bf n}'$ satisfy an {\em orthogonality} relation:
\be
\frac{(2s+1)}{4\pi} \mbox{ Tr } \left[ \, {\widehat \Delta} ({\bf n}) \,  
                     {\widehat \Delta} ({\bf n}') \, \right] 
                 = \delta ({\bf n} - {\bf n}') \, , 
                                   \qquad {\bf n}, {\bf n}' \in {\cal S}^2 \, 
\label{ortho}
\ee
This property, also called {\em traciality}, allows one to invert (\ref{definesymbol}):
\be 
{\widehat A} = \frac{(2s+1)}{4\pi} \int_{{\cal S}^2} d{\bf n} \, W_{\! \! A} ({\bf n})  
\, {\widehat \Delta} ({\bf n})  \, ,
\label{invertdefinesymbol}
\ee
that is, to express the operator ${\widehat A}$ in terms of its symbol. Note that (\ref{definesymbol}) and (\ref{invertdefinesymbol}) are effected by means of the {\em same} operator kernel.}
\item[($\delta$):]{The kernels transform {\em covariantly} with respect to the group $SU(2)$:
\be
 {\widehat U}_{\sf R}\, {\widehat \Delta ({\bf n})} \,  {\widehat U}_{\sf R}^\dagger 
= {\widehat \Delta ( {\sf R} \, {\bf n})} 
\label{covar}
\ee
where the unitary ${\widehat U}_{\sf R}$ in ${\cal H}_s$ represents a rotation ${\sf R}$ .}
\end{enumerate}
These conditions have been used to derive the explicit form of the 
kernels ${\widehat \Delta ({\bf n})}$ for a spin. For a quantum {\em particle} in one dimension, the family of kernels generating the Moyal representation is given by ${\widehat \Delta} (\alpha) = {\widehat T} (\alpha) \, {\widehat P}\,  {\widehat T}^\dagger (\alpha)$, where ${\widehat T} (\alpha)$, $\alpha \in {\sf C}$, represents a phase-space translation on the particle Hilbert space, and ${\widehat P}$ is the parity operator. Upon replacing the group of rotations by the translations in phase space, the operator ${\widehat \Delta} (\alpha)$ satisfies properties structurally equal to ($\alpha$)-($\delta$), giving thus rise to the standard phase-space representation of quantum mechanics for a particle. 

In the following, the properties of the expectation-value representation will be compared to those of the Stratonovich-Weyl correspondence. A fundamental difference is obvious from the outset: the representation based on the quorum $\cal Q$ and its dual 
${\cal Q}^D$ gives rise to {\em two} intimately connected types of symbols, $A^n$ and ${\sf A}_n$. This `doubling' is due to fact that,individually, the collections $\{ {\widehat Q}_n \}$ and $\{ {\widehat {\sf Q}}^n\} $ do not form two orthogonal bases but only together they define a {\em bi-orthogonal} basis of the space ${\widehat {\cal A}}_s$. 

The comparison  will be simplified by slightly modifying the notation. Denote the symbols of ${\widehat A}$ in analogy to (\ref{definesymbol}) by 
\be
V_{\!\!A}^{n} \equiv A^{n} 
   = \mbox{ Tr } \left[ {\widehat {A}} \, {\widehat {\sf Q}}^{n} \right] 
                             \quad \mbox{ and } \quad
{\sf V}_{\!\!A,n} \equiv {\sf A}_{n}  
   = \mbox{ Tr } \left[ {\widehat {A}} \, {\widehat Q}_{n} \right] \, .
\label{definesymbol2}
\ee
The operator kernels ${\widehat Q}_n$ and ${\widehat {\sf Q}}^n$ have the following properties:
\begin{enumerate}
\item[($\alpha'$):]{Each of the kernels ${\widehat Q}_n$ and ${\widehat {\sf Q}}^n$ is a hermitean operator,
\be
{\widehat Q}_n^\dagger  =  {\widehat Q}_n  
           \quad \mbox{ and } \quad
 \left( {\widehat {\sf Q}}^n \right)^\dagger  =  {\widehat {\sf Q}}^n \, , \qquad 
                         n=1,\ldots , N_s \, , 
\label{herm2}
\ee
implying that the symbols $V_{\! \! A,n}$ and $V_{\!\!A}^{n}$ of a hermitean operator ${\widehat A}$ are {\em real}.}
\item[($\beta'$):]{Both sets of kernels are {\em complete}:
\be
\frac{1}{2s+1} \sum_{n=1}^{N_s} 
      \mbox{ Tr} \left[ \, {\widehat Q}_n \, \right] \, {\widehat {\sf Q}}^n   
  = 
\frac{1}{2s+1} \sum_{n=1}^{N_s} 
      \mbox{ Tr } \left[ \, {\widehat {\sf Q}}^n \, \right] \, {\widehat Q}_n   
  = {\widehat E} \, ,
\label{compl2}
\ee
providing thus two explicit resolutions of the identity. The analogy to ($\beta$) 
becomes more obvious if one notes that the kernel in (\ref{compl}) is multiplied 
by the symbol of the unity, $W_E ({\bf n}) \equiv \mbox{ Tr } [ {\widehat E} {\widehat \Delta} ({\bf n}) ] = 1$, and the coefficients in (\ref{compl2}) are the symbols of the unity, too: ${\sf V}_{E,n} \equiv \mbox{ Tr } [ {\widehat E} \, {\widehat Q}_n  ]$ and $V_E^n \equiv \mbox{ Tr } [ {\widehat E} \, {\widehat {\sf Q}}^n ]$.} 
\item[($\gamma'$):]{Two kernels with labels $n$ and $n'$ satisfy the {\em bi-orthogonality} relation (\ref{orthogonalityquorum}) in the space ${\cal H}_s$:
\be
\frac{1}{2s+1} \mbox{ Tr }\left[  {\widehat Q}_{n} {\widehat {\sf Q}}^{n'} \right] 
= \delta_{n}^{n'}  \, , 
\qquad 1 \leq n,n' \leq N_s \, .
\label{ortho2}
\ee
This property allows one to invert (\ref{definesymbol2}):
\be 
{\widehat A} 
   = \frac{1}{2s+1} \sum_{n=1}^{N_s} {\sf V}_{\!\!A,n} {\widehat {\sf Q}}^n   
   = \frac{1}{2s+1} \sum_{n=1}^{N_s} V_{\!\!A}^{n} \, {\widehat Q}_n   
   \, ,
\label{invertdefinesymbol2}
\ee
that is, to express the operator ${\widehat A}$ in terms of its symbols. Contrary to  (\ref{definesymbol}) and (\ref{invertdefinesymbol}) the transformations (\ref{definesymbol2}) and (\ref{invertdefinesymbol2}) are not effected by means of the same but the {\em dual} operator kernel.}
\item[($\delta'$):]{The kernels ${\widehat Q}_{n}$ transform {\em covariantly} with respect to the group $SU(2)$: 
\be
{\widehat U}_{\sf R} \, {\widehat Q}_{n} \, {\widehat U}_{\sf R}^\dagger 
 = {\widehat U}_{\sf R} \, \ket{ {\bf n}_n} \bra{{\bf n}_n} \, {\widehat U}_{\sf R}^\dagger 
 = \ket{{\sf R}\, {\bf n}_{n}} \bra{{\sf R}\, {\bf n}_{n}} = {\widehat Q}_{{\sf R},n}\, ,
\label{covar2}
\ee
since a rotation $\sf R$ maps a coherent state with label ${\bf n}$ to another coherent state with label $\sf R \, {\bf n}$. Hence, the operators ${\widehat Q}_{n}$ transform
covariantly, and a similar relation holds for the dual family.}   
\end{enumerate}
Conceptually, the quorum ${\cal Q}$ and its dual ${\cal Q}^D$ do thus provide a generalization of the Stratonovich-Weyl correspondence which is based on a bi-orthogonal basis of the space of operators ${\cal A}_s$. Therefore, this formulation differs also from the Wigner formalism for Hilbert spaces of finite systems introduced in Ref.\  \cite{leonhardt95}.   
\section{Outlook}
It has been shown that state reconstruction for a spin based on coherent states gives rise to a generalized Stratonovich-Wey correspondence mapping operators on phase-space functions. In a natural manner, {\em two} symbols (each a collections of $N_s$ numbers 
associated with specific points of phase space) with qualitatively different properties emerge. One of the symbols allows one to represent a density matrix in terms {\em positive} numbers which correspond to probabilities associated with incompatible measurements.

\medskip

\noindent {\bf Acknowledgments}

Financial support by the {\em Schweizerische Nationalfonds} is gratefully acknowledged.  
\end{document}